# Nearest-neigbor spacing distributions of the $\beta$-Hermite ensemble of random matrices

G. Le Caër [a], C. Male [b] and R. Delannay

Groupe Matière Condensée et Matériaux, C.N.R.S. U.M.R. 6626, Université de Rennes-I, Campus de Beaulieu, Bât. 11A, Avenue du Général Leclerc, F-35042 Rennes Cedex, France

[a]corresponding author, E-mail : gerard.le-caer@ univ-rennes1.fr

[b] permanent address : Ecole Normale Supérieure de Cachan, Campus de Kerlann, F-35170 Bruz, France

## Abstract

The evolution with $\beta$ of the distributions of the spacing '$s$' between nearest-neighbor levels of unfolded spectra of random matrices from the $\beta$-Hermite ensemble ($\beta$-HE) is investigated by Monte Carlo simulations. The random matrices from the $\beta$-HE are real-symmetric and tridiagonal where $\beta$, which can take any positive value, is the reciprocal of the temperature in the classical electrostatic interpretation of eigenvalues. The distribution of eigenvalues coincide with those of the three classical Gaussian ensembles for $\beta$=1, 2, 4. The use of the $\beta$–Hermite ensemble results in an incomparable speed up and efficiency of numerical simulations of all spectral characteristics of large random matrices. Generalized gamma distributions are shown to be excellent approximations of the nearest-neighbor spacing (NNS) distributions for any $\beta$ while being still simple. They account both for the level repulsion in : $s^\beta$ when $s \to 0$ and for the whole shape of the NNS distributions in the range of '$s$' which is accessible to experiment or to most numerical simulations. The exact NNS distribution of the GOE $(\beta = 1)$ is in particular significantly better described by a generalized gamma distribution than it is by the Wigner surmise while the best generalized gamma approximation coincides essentially with the Wigner surmise for $\beta > \sim 2$. They describe too the evolution of the level repulsion between that of a Poisson distribution and that of a GOE distribution when $\beta$ increases from 0 to 1. The distribution of $\ln(s)$, related to the electrostatic interaction energy between neighboring charges, is accordingly well approximated by a generalized Gumbel distribution for any $\beta \geq 0$. The distributions of the *minimum* NN spacing between eigenvalues of matrices from the $\beta$-HE, obtained both from as-calculated eigenvalues and from unfolded eigenvalues are Brody distributions which are classically used to characterize the spectral fluctuations of various physical systems.

# 1. Introduction

Random matrix theory (RMT) continues to be important in various branches of physics as different as quantum chaology, for investigating growth models or in econophysics to quote just a few [1-8]. The asymptotic distribution of the spacing '$s$' between consecutive energy levels of quantum systems or between successive eigenvalues of random matrices or points of 1D point processes, $p(s)$, called most often nearest-neighbor spacing (NNS) distribution, is frequently investigated in statistical physics. The considered distribution of adequately processed NN spacings is compared to those of spacings between successive eigenvalues, once 'unfolded' (section 3), of reference ensembles of $N \times N$ random matrices for $N$ large. In general, these spacings are rescaled so that their average is $\langle s \rangle = 1$, where $\langle f(s) \rangle = \int_0^\infty f(s) p(s) ds$. Theoretical distributions are rarely available, experimental or simulated distributions are used instead and compared to approximations of the reference asymptotic distributions. Many characteristics of the distributions of eigenvalues of $N \times N$ random matrices from the three fundamental Gaussian ensembles, the GXE's where X= O, U, S means orthogonal, unitary and symplectic respectively, are indeed known both exactly at finite $N$ and asymptotically at large $N$ for which local statistics are often universally distributed once properly scaled. Matrices of the three fundamental Gaussian ensembles are real symmetric for the GOE, Hermitian for the GUE and quaternion self-dual for the GSE. A supplementary ensemble, the Gaussian diagonal ensemble (GDE), is made from matrices whose sole non-zero elements are diagonal with identical and independent normal distributions.

Without aiming at completeness, we briefly depict contributions which NNS distributions have made to gain a better knowledge of a variety of topics in statistical physics. Such distributions have been thoroughly investigated in quantum chaology [2-6, 8]. The level fluctuations of quantum Sinai's billiards were early recognized to be consistent with those of the GOE eigenvalues ([6], 1984). Semiclassical NNS distributions, notably the classical Berry-Robnik NNS distribution, were derived for systems whose classical energy surface is divided into separate regions in which motion is regular or chaotic [8]. Most often, phenomenological models of the evolution of the NNS distributions are used to describe specific transitions between the Wigner-Dyson and the Poisson statistics. A recent example is that of the Poisson-to-Wigner crossover transition of the NNS distribution of random points on fractals as modelled by a Brody distribution [9]. The Wigner surmise in its most general form approximates the spacing distribution between successive eigenvalues in the bulk of the unfolded spectra of various random matrices:

$$\begin{cases} p_{W,\beta}(s) = a_W s^\beta \exp(-b_W s^2) \\ a_W = \dfrac{2\left[\Gamma((2+\beta)/2)\right]^{\beta+1}}{\left[\Gamma((1+\beta)/2)\right]^{\beta+2}} \qquad b_W = \left(\dfrac{\Gamma((2+\beta)/2)}{\Gamma((1+\beta)/2)}\right)^2 \end{cases} \qquad (1)$$

($\langle s \rangle = 1$). This celebrated Wigner distribution (eq. 1) approximates well the NNS distributions of the GOE for $\beta=1$, and very well those of the GUE and of the GSE for $\beta=2$, 4 respectively [1]. They define in particular three universality classes of the level repulsion at small '$s$': $s \to 0$, $p_{W,\beta}(s) \sim s^\beta$. The Wigner surmise is further discussed in section 4 and in appendix A for the $\beta$-Hermite ensemble investigated here. The asymptotic NNS distribution of the Gaussian diagonal ensemble ($\beta=0$) is a Poisson distribution as independently and identically distributed (iid) diagonal elements once unfolded are simply iid random variables uniformly distributed between 0 and 1. Wigner surmises may be derived too for the spacing between two levels separated by $k$ levels as done for instance for chaotic systems although they are not nearly as good as that for NNS [10]. The phenomenological Brody distribution [2], which is in common use, will be considered in greater detail. It depends on a unique parameter, denoted hereafter as $\omega_B$:

$$\begin{cases} p_{B,\omega_B}(s) = c_B(1+\omega_B)s^{\omega_B} \times \exp\{-c_B s^{1+\omega_B}\} \\ c_B = \left[\Gamma\left((2+\omega_B)/(1+\omega_B)\right)\right]^{1+\omega_B} \end{cases} \quad (2)$$

The Brody distribution interpolates between the NNS distribution of the GOE as approximated by the Wigner surmise ($\omega_B=\beta=1$) and a Poisson distribution ($\omega_B=0$). The one-parameter Izrailev's distribution is sometimes relevant as a phenomenological NNS distribution (see for instance, Varga et al. [11] and Molina et al. [12]) but it is not as easily handled as is the Brody distribution.

Among the NNS distributions studied in physics, let us further mention those of disordered systems near the metal-insulator transition [11], of transfer matrices of classical lattice spin models (2D and 3D Ising models, Potts model) [13] and finally those of unfolded eigenvalues of the matrix of the second derivatives of the potential energy for a classical atomic liquid [14] and for a supercooled liquid [15], in relation with instantaneous normal vibration modes. In random matrix theory, the exact asymptotic NNS distribution, with an $\exp(-as^{8/3})$ term, was derived for random Hermitian matrices in an external source [16] and the Wigner surmises were calculated for random matrix ensembles which maximise a nonextensive entropy [17]. The effect of removing levels at random from spectra of the classical gaussian ensembles was recently characterized [18]. Renewed efforts were further directed to the exact asymptotic NNS distributions of the Gaussian ensembles and to the associated Gaussian fluctutations [19-20]. Forrester and Witte [19] expressed the exact NNS distributions of the GOE and the GUE in the form of a Wigner surmise, $a(s)\exp(-b(s))$, where $a(s)$ and $b(s)$ are deduced from Painlevé transcendents. Of interest is the study of the terrace width distribution whose connection with the classical NNS distributions of random matrix theory helps to extract the strength of the interaction between steps on a vicinal surface [21]. More remote at first sight are studies that are nevertheless relevant to statistical physics as those of networks [22], of the bus spacing distribution in a mexican city with its description by the NNS distribution of the GUE and its recent modelling by Baik et al. [23], of the gap distribution of parked cars [24] and of the gene coexpressions [25]. The repulsion effect between eigenvalues of asymmetric complex random matrices was curiously found to account well for the 2D distribution of the district chieftowns of mainland France [26]. Finally, notable is the investigation of the spacing distribution of the imaginary parts of the zeros of the Riemann zeta function and of a variety of their characteristics which all link number theory and RMT [4,27].

At that point, the properties of eigenvalues of Gaussian ensembles are recalled to be interpreted in 2D from the equilibrium characteristics at a temperature $1/\beta$ of a gas of $N$ identical point charges on a line [1], often referred to as a log-gas [28], which interact via a logarithmic Coulomb potential and are confined by an external harmonic potential. A newcomer in RMT, the $\beta$-Hermite ensemble ($\beta$-HE) of tridiagonal random matrices [29] (section 2), whose temperature $1/\beta$ can take any value, offers an enlarged perspective. Its joint distribution of eigenvalues is identical with those of the previous Gaussian ensembles for $\beta$=1, 2, 4 respectively [29-33]. The $\beta$-HE owes its name to the Hermite orthogonal polynomials whose weight $\propto \exp(-ax^2)$ is associated with the harmonic potential. Among others, the Hermite polynomials make it possible to derive closed form expressions of the correlation functions of all orders for Gaussian ensembles [1]. The use of $\beta$-H matrices results, among others, in an unrivalled speed-up of numerical simulations of all characteristics of the eigenvalue distributions of large random matrices (section 3).

When exact asymptotic NNS distributions are not available or are out of reach, a situation which is the rule, it is necessary to resort to approximate distributions. The latter must be simple as are the Wigner surmise and the Brody distribution while hopefully improving their effectivities and giving the possibility to test the physical relevance of the fitted parameters. In the present paper, we shall take advantage of the flexibility of the $\beta$-Hermite ensemble and of its numerical efficiency to show that the use of a NNS distribution which depends on two shape parameters, namely the generalized gamma (GG) distribution, allows us to give a clear meaning to one of them. An *a posteriori* one-parameter approximation of the NNS distributions of the $\beta$-Hermite ensemble shall then be obtained for any $\beta$. The latter improves notably the usual approximation to the NNS distribution of the Gaussian orthogonal ensemble in the range of spacing which can be accessessed experimentally or by computer simulation. Advantage shall be taken of the fact that a Brody distribution is nothing else than a Weibull distribution of extreme value to investigate the distribution of the minimum nearest-neighbor spacing of the $\beta$-Hermite ensemble.

## 2. The $\beta$–Hermite ensemble of Dumitriu and Edelman [29]

The number of distinct real random variables, which are necessary to construct a $N \times N$ GXE matrix, is:

$$N_\beta = N + \beta \frac{N(N-1)}{2} \qquad (3)$$

with $\beta$=0, 1, 2, 4 for the GDE, the GOE, the GUE and the GSE respectively. The joint distribution of eigenvalues $(\lambda_1, \lambda_2,..,\lambda_N)$ of $N \times N$ random matrices from the Gaussian ensembles is [1]:

$$\begin{cases} P_{N,\beta}(\lambda_1,..,\lambda_N) = K_{N,\beta} \exp\left(-\frac{1}{2\sigma^2}\left[\sum_{k=1}^{N}\lambda_k^2\right]\right)\left[\prod_{1 \leq j < k \leq N}|\lambda_j - \lambda_k|^\beta\right] \\ \rho = \frac{\beta}{2} \qquad K_{N,\beta} = \sigma^{-N_\beta/2}(2\pi)^{-N/2}\prod_{j=1}^{N}\frac{\Gamma(1+\rho)}{\Gamma(1+j\rho)} \end{cases} \qquad (4)$$

where $K_{N,\beta}$ is the reciprocal of the Mehta integral ([1] p.354). The $N_\beta$ distinct elements of the GXE matrices are recalled to be independently distributed according to Gauss distributions with zero means and variances $\sigma_{ij}^2 = \sigma^2(1+\delta_{ij})/2$, denoted hereafter $N(0,\sigma_{ij}^2)$ $\left(g_{ij}(x) = \exp(-x^2/2\sigma_{ij}^2)/\sigma_{ij}\sqrt{2\pi}\right)$.

Dumitriu and Edelman [29] derived the $\beta$-Hermite ensemble of real-symmetric tridiagonal random matrices whose eigenvalue density is given by eq.4 for any $\beta \geq 0$ [29-32]. Extensions of other classical ensembles were similarly performed, for the $\beta$-Laguerre ensemble among others [29]. A $N \times N$ random matrix from the $\beta$-HE is defined as:

$$\mathbf{A}_{N,\beta} = \sigma \mathbf{H}_{N,\beta} = \sigma \begin{bmatrix} H_{11} & H_{12}/\sqrt{2} & 0 & . & 0 \\ H_{12}/\sqrt{2} & H_{22} & H_{23}/\sqrt{2} & 0 & . \\ 0 & H_{23}/\sqrt{2} & . & . & 0 \\ . & 0 & . & H_{N-1,N-1} & H_{N-1,N}/\sqrt{2} \\ 0 & . & 0 & H_{N-1,N}/\sqrt{2} & H_{NN} \end{bmatrix} \qquad (5)$$

The $2N-1$ distinct matrix elements, namely $A_{kk} = \sigma H_{kk}$ $(k=1,...,N)$ and $A_{k,k+1} = \sigma H_{k,k+1}/\sqrt{2}$ $(k=1,...,N-1)$ are independently but not identically distributed and $\sigma$ is a scale factor. Every $H_{kk}$ has a $N(0,1)$ Gauss distribution while the off-diagonal element $x_k = H_{k,k+1}$ $(k=1,...,N-1)$ has a chi distribution with $k\beta$ degrees of freedom whose probability density is :

$$q_{N,\beta}(x_k) = 2^{1-k\beta/2} x_k^{k\beta-1} \exp\left(-x_k^2/2\right) \Big/ \Gamma(k\beta/2) \qquad (x_k \geq 0) \qquad (6)$$

It follows immediately that $tr(H_{N,\beta}^2)$, a sum of independent chi-squared random variables, has a chi-squared distribution with $N_\beta$ degrees of freedom for any $\beta$ (eq. A-3).

When $\beta \to \infty$, the off-diagonal element $H_{k,k+1}$ can be written as $\sqrt{k\beta} + X/\sqrt{2}$, where $X$ is a standard Gaussian [29]. Then, the properly rescaled eigenvalues, whose large $\beta$ distribution tends to a multivariate Gaussian distribution, fluctuate around the $N$ roots of the $N$th Hermite polynomial. These fluctuations around the equilibrium positions are correlated in a rather involved manner [30]. Andersen et al. [34] derived the multivariate Gaussian distribution of the eigenvalues by expanding the logarithm of the multivariate probability given by eq.4 (with $\sigma^2 = 1/\beta N$) in the vicinity of its maximum to describe the normal modes of the eigenvalue spectrum. The most probable fluctuation in the spectrum corresponds to a common shift of all eigenvalues, without change in their relative separation, which reflects the spectral rigidity [34]. Andersen et al. showed that the next most probable mode is a breathing mode.

## 3. Computer simulations

The $\beta$-Hermite ensemble is particularly suited for efficient numerical simulations of its various characteristics with computer times essentially independent of $\beta$. We performed Monte Carlo calculations in Fortran with a standard laptop computer to simulate $\beta$-H matrices (times of 0.03s and 0.12s were for instance needed to build and to diagonalize a 200x200 matrix and a 400x400 matrix respectively). Gaussian variables were generated by the polar Box-Muller method [35]. The chi distributions of the non-diagonal elements were generated through gamma distributions, $G_{a,b}(x) = x^{a-1}\exp(-x/b)/(\Gamma(a)b^a)$, using the Johnk's generator for parameters $a<1$ ([35] p.418) and the Best rejection algorithm for $a > 1$ ([35] p.410). We notice in passing that $(-\ln U)^{1/(1+\omega_B)} \Big/ \Gamma\left(\dfrac{2+\omega_B}{1+\omega_B}\right)$ is Brody distributed (eq.2) when $U$ is uniformly distributed between 0 and 1.

Once generated and diagonalized, the eigenvalue spectrum of a $\beta$-Hermite matrix must be 'unfolded' to derive the sought-after NN spacings. The unfolding process was performed from the Wigner semi-circle of radius 1:

$$\rho_W(\lambda) = \frac{2}{\pi}\sqrt{(1-\lambda^2)} \qquad (7)$$

for $|\lambda| \leq 1$ and 0 elsewhere. A scale parameter, $\sigma = \dfrac{1}{\sqrt{4+2\beta(N-1)}}$ (eq. 4), gives $\langle \lambda^2 \rangle = \dfrac{1}{4}$ and thus a semi-circle of radius 1. Any eigenvalue $\lambda_k$, which belongs to $(-r, +r)$, with typical values of $r(<1)$ ranging between 0.8 and 0.9, is transformed into an unfolded eigenvalue $\lambda_k^{(u)}$, which is the value of the cumulative distribution function, $F(\lambda) = \int_{-\infty}^{\lambda} \rho(x) dx$ of the smoothed level density $\rho(x)$ for $\lambda = \lambda_k$. The cumulative distribution function is either calculated exactly or estimated numerically. When the eigenvalue density differs negligibly from a Wigner semi-circle in the selected range, the unfolding process is then performed as follows:

$$\lambda_k^{(u)} = \int_{-1}^{\lambda_k} \rho_W(\lambda) d\lambda = \frac{1}{2} + \frac{\lambda_k \sqrt{1-\lambda_k^2}}{\pi} + \frac{\sin^{-1}\lambda_k}{\pi} \qquad (8)$$

The unfolded density of eigenvalues is constant by construction. The unfolded eigenvalues are finally rescaled so that the average spacing between nearest neighbors is $\langle s \rangle = 1$. Eq. 8 requires thus that the eigenvalue density is, to an excellent approximation, a Wigner semi-circle in $(-r, +r)$. This is the case for moderate values of $N$ (some tens) when $\beta$ ranges between ~1 and ~5 while deviations occur both for low and for high values of $\beta$. At a high temperature $1/\beta$ ($\beta = 1$), $N$ must be increased as the density, which is intermediate between a Gaussian shape and a Wigner semi-circle for moderate matrix sizes, evolves towards the Wigner semi-circle for large matrix sizes (figures 1a and 2). At low temperature, the progressive freezing of charges around their equilibrium positions produces oscillations of the eigenvalue density around the smooth Wigner semi-circle [31-33] (figure 1b). The oscillations seen in figure 1b are not due to random fluctuations but are due to the real shape of the eigenvalue density. When binned, its appearance is related to the interplay between the local wavelength of the oscillations and the bin size which is here 0.01. Again, the matrix size must be significantly increased to damp such oscillations. When the empirical cumulative distribution shows significant local deviations from a Wigner semicircle, the unfolding process was performed from a smooth eigenvalue density obtained numerically as the average of an ensemble of spectra simulated with Matlab. Another method, sometimes used, would be to retain just one spacing between central raw eigenvalues. We verified for some values of $\beta$ that the latter method and the present method give the same results as a consequence of ergodicity.

For 0.05≤β≤5, the simulated distributions shown in the present paper were obtained altogether from simulations of $10^5$ matrices with $N$=100, of $5.10^4$ matrices with $N$=200, of $10^3$ matrices with $N$=500 and of 100 matrices with $N$=5000.

Matrices of size $N$=60 from the β-HE, typically $10^6$, were generally used to calculate the distributions of the minimum NN spacing $s_{min}$ as a function of $β$ both from the as-calculated eigenvalues $\lambda_k$ and from the unfolded eigenvalues $\lambda_k^{(u)}$. The position of $s_{min}$ calculated from the $\lambda_k^{(u)}$'s fluctuates in the whole interval associated with the selected range $(-r,+r)$ of the $\lambda_k$'s, essentially without systematic trends when $β <$~4. A range (-1,+1) was retained for calculating $s_{min}$ from the $\lambda_k$'s. Another method would be to keep solely the central spacing of every $N \times N$ matrix and to determine the distribution of an ensemble of $s_{min}$, each being the minimum of $k$ central spacings. In the following, $p_β(x)$ $(x = s$ or $x = s_{min})$ denotes equally the spacing distribution obtained by Monte Carlo simulations of β-Hermite random matrices and the exact asymptotic distribution at the temperature $1/β$.

## 4. Approximations of the simulated NNS distributions of the β-HE

### 4.1. The spacing distribution for N=2

The level spacing distributions of large matrices are well approximated by those of ensembles of $N$=2 matrices (see for instance, Haake [8] and refs 36-37). The Wigner surmise (eq. 1) is derived in a simple way in appendix A for the β-HE. It is exact, by construction for any $β$, for the spacing distribution of a $2 \times 2$ β-H matrix. The latter method is however not necessarily valid for any ensemble of random matrices. We construct for instance in appendix B a 'double-chi' ensemble of real-symmetric matrices whose distinct off-diagonal elements depend on a single parameter $β$. It has a level repulsion, $s \to 0$, $p(s) \sim s^β$, which is that of the β-HE for any $β$ for $N$=2 but its asymptotic level repulsion is linear at large $N$, $s \to 0$, $p(s) \sim s$, for any $β$>0 (appendix B and figure 13). The method works well for the Gaussian ensembles as the Wigner surmise is a fair approximation of the exact asymptotic NNS densities of the GXE's. The deviations between the exact NNS distribution of the GOE and the Wigner surmise are further discussed in section 4.3.

### 4.2. The NNS distributions of the β–HE and the generalized gamma distribution

The Brody distribution (eq.2) is used very frequently in the literature as a semi-empirical model of the evolution of NNS distributions between a Wigner distribution and a Poisson distribution. It is a convenient distribution with a sole free parameter $\omega_B$ but its functional form is curiously considered somewhat as unmodifiable despite its moderate success in accounting for some NNS distributions. The Brody distribution, the Wigner surmise and the Poisson distribution belong all to the same family, namely that of the generalized gamma distribution (GG) [38,43] which reads:

$$\begin{cases} p_{\omega_1,\omega_2}(s) = a_{\omega_1,\omega_2} s^{\omega_1} \exp\left(-b_{\omega_1,\omega_2} s^{\omega_2}\right) \\ a_{\omega_1,\omega_2} = \frac{\omega_2 \left[\Gamma\left((2+\omega_1)/\omega_2\right)\right]^{\omega_1+1}}{\left[\Gamma\left((1+\omega_1)/\omega_2\right)\right]^{\omega_1+2}} \quad b_{\omega_1,\omega_2}(s) = \left(\frac{\Gamma\left((2+\omega_1)/\omega_2\right)}{\Gamma\left((1+\omega_1)/\omega_2\right)}\right)^{\omega_2} \end{cases} \quad (9)$$

when the average spacing is $\langle s \rangle = 1$. It is thus quite natural to try to approximate all simulated NNS distributions by a GG distribution without *a priori* constraints on the shape parameters $\omega_1, \omega_2$. The raw moments of this GG distribution are:

$$\langle s^n \rangle = \frac{\left[\Gamma\left(\frac{1+\omega_1}{\omega_2}\right)\right]^{n-1}}{\left[\Gamma\left(\frac{2+\omega_1}{\omega_2}\right)\right]^n} \times \Gamma\left(\frac{1+n+\omega_1}{\omega_2}\right) \quad (10)$$

The variance is thus:

$$\text{var}(s) = \langle (s-1)^2 \rangle = \frac{\Gamma\left(\frac{1+\omega_1}{\omega_2}\right)\Gamma\left(\frac{3+\omega_1}{\omega_2}\right)}{\left[\Gamma\left(\frac{2+\omega_1}{\omega_2}\right)\right]^2} - 1 \quad (11)$$

Distributions belonging to the GG family were sometimes derived theoretically for specific models of NNS distributions, for instance for the metal-insulator transition of the Anderson model (Varga et al. in ref. 11). The distributions of the NN spacing between eigenvalues of adjacency matrices of various random networks were investigated very recently by Palla and Vattay [22]. These distributions are sensitive to a percolation transition which occurs when the average valence of the graph vertices is 1. NNS distributions were found to be well described by a Brody distribution and by a gamma distribution ($\omega_2 = 1$) above and below the critical point respectively [22].

However, the GG distribution with two free parameters, $\omega_1$ and $\omega_2$, is seemingly not considered *per se* as an appropriate and robust tool for approximating NNS distributions in broad conditions. It might appear that this supplementary degree of freedom is nothing else than the price to pay for a somewhat artificial improvement of the fit of the real distribution by an approximate one. As shown below, this approach is yet beneficial as the freeness of the parameter $\omega_2$ permits to recover the real physical meaning of $\omega_1$ in the whole temperature range.

As exemplified by figure 3, the Brody distribution is inadequate to model the NNS distributions of the $\beta$-HE except around $\beta=1$. The Brody parameter $\omega_B$ is not related to $\beta$ in a simple way as shown by figure 4 except in the range (0-1) (figure 4a) where it is an approximation of $\beta$ not as accurate as is $\omega_1$ (typically $\beta \pm$ some $10^{-3}$).

The shape parameters $\omega_1$ and $\omega_2$, obtained by least-squares fitting the simulated NNS distributions $p_\beta(s)$ by GG distributions, are shown as a function of $\beta$ in figures 4a, 5 and 6. The GG distributions describe very well the simulated distributions for any $\beta$ as shown by figures 7 and 8. The particular case of the GOE ($\beta=1$) is considered more particularly in section 4.3.

The parameter $\omega_1$ is, to a very good precision, equal to $\beta$ (figures 4a and 5) whatever $\beta$ while the shape parameter $\omega_2 = 2 - \omega$ increases monotonically from 1 to 2 when $\beta$ increases from 0 to $\infty$. The deviation to 2, $\omega$, is very well approximated by a stretched exponential (figure 6). To summarize:

$$\omega_1 = \beta, \quad \omega_2 = 2 - \omega = 2 - \exp(-2.12\beta^{0.75}) \qquad (12)$$

($2.12 \pm 0.08$ and $0.75 \pm 0.03$). The level repulsion for small spacings varies thus as $s^\beta$, for any $\beta > 0$, as expected from the Wigner surmise $p_{W,\beta}(s)$ (eq. 1). The GG distribution reduces essentially to the Wigner surmise when $\beta \gtrsim \approx 2$ (table 1) while it differs from it and improves its accuracy in modelling the distributions $p_\beta(s)$ for $\beta \leq 2$. Equations 9 and 12 yield a Gaussian distribution when $\beta \to \infty$ (eq. A-8 and figure 8).

An *a posteriori* approximate NNS distribution of the $\beta$-HE reads finally ($\langle s \rangle = 1$):

$$\begin{cases} p_{\beta,\omega}(s) = a_{\beta,\omega} s^\beta \exp\left(-b_{\beta,\omega} s^{2-\omega}\right) \qquad \omega_2 = 2 - \omega \\ \\ a_{\beta,\omega} = \dfrac{\omega_2 \left[\Gamma((2+\beta)/\omega_2)\right]^{\beta+1}}{\left[\Gamma((1+\beta)/\omega_2)\right]^{\beta+2}} \qquad b_{\beta,\omega} = \left(\dfrac{\Gamma((2+\beta)/\omega_2)}{\Gamma((1+\beta)/\omega_2)}\right)^{\omega_2} \end{cases} \qquad (13)$$

for any $\beta \geq 0$, $\omega_1$ was actually taken as $\omega_1 = \beta$ and only $\omega_2$ was obtained from the NNS distributions $p_\beta(s)$ shown in figure 7. As $\omega_2$ is very close to 2 for $\beta = 2,4$ (table 1), we restrict the following discussion to the particular case $\beta = 1$.

### 4.3. The particular case of the GOE

When the exact asymptotic spacing distribution of the GOE ([1] and Haake [8]) is least-squares fitted by a Brody distribution (eq.2), $\omega_B$ differs from the value of 1 it should have if $p_{B,1}(s) \equiv p_{W,1}(s)$, being actually $\omega_B = 0.958(3)$. That difference is indeed the sole way to reduce partly the deviations which exist between the Wigner surmise $p_{W,1}(s)$ and the exact asymptotic distribution $p_1(s)$ [28] (figure 9, see also Haake [8] figure 4.2 p. 55). The GOE linear level repulsion $s \to 0$, $p(s) \sim s$ is then only approximately accounted for by $p_{B,0.958}(s)$. The GG distribution $p_{1,\omega_2}(s)$ ($\omega_2 = 1.886$) (eq.13) describes better the whole shape of $p_1(s)$ than do the previous distributions (figure 9a). That point is confirmed by Monte Carlo simulations (figure 9b). Further, the regular fluctuations of MCS-W around the exact E-W curve are an indication of the correctness of the simulated distributions and give further confidence in the improvement brought about by the GG distribution. The moments $\langle s^n \rangle$ ($n = 2,3,4$), the coefficient of skewness and the coefficient of excess of $p_{1,\omega_2}(s)$ are compared to the exact moments in table 2. Even if the large-$s$ asymptotics of the previous GG distributions are not those expected for the exact NNS distributions ([1,8,44] and [20] for the GUE), we conclude that $p_{1,1.886}(s)$ is a better, still simple, approximation of $p_1(s)$ than is the Wigner surmise $p_{W,1}(s)$ in the whole range of '$s$' which is accessible to experiment or to most numerical simulations. A convenient, still satisfying, mnemonic NNS distribution for the GOE, is finally expressed as:

$$p_{1,\omega_2}(s) = a_1 s \exp\left(-b_1 s^{2-1/9}\right) \qquad (14)$$

Eqs 13 and 14 have a 'Wigner surmise' form, $a(s)\exp(-b(s))$, as does the exact NNS distribution for the GOE in which both $a(s)$ and $b(s)$ depend on Painlevé transcendents (Forrester and Witte [19]). Although the GG distribution accounts very well for the whole shape of the NNS distribution in the range of '$s$' accessible in practice, it is approximate as $a(s) \neq cst \times s$ and $b(s) \neq cst \times s^\alpha$ [19].

To conclude, GG distributions (eq.13) are excellent approximations of the asymptotic NNS distributions of the $\beta$-HE for any $\beta$. For large $\beta$, it is essentially identical with the Wigner surmise while it is much better than the Brody distribution for $0 \leq \beta \leq 1$. Equivalently, the distribution of $x = \ln(s)$ is well approximated by a generalized Gumbel distribution.

**4.4. The relation between a GG distribution and a generalized Gumbel distribution.**

Large scale fluctuations in many correlated systems are described by the generalized Gumbel distribution [45]. It is further the asymptotic distributions of the extreme values of sequences correlated over a 'distance' 1/a (eq.15) [46] and the asymptotic distribution of sums of independent non identically distributed random variables or of correlated variables which do not verify the conditions of validity of the central limit theorem [47]. The generalized Gumbel distribution reads:

$$g_{a,b,m}(x) = \frac{a^a}{|b|\Gamma(a)} \exp\left(a\left\{\frac{x-m}{b} - \exp\left(\frac{x-m}{b}\right)\right\}\right) \quad (15)$$

where $a > 0$ and $b > 0$ in most cases studied below. The characteristic function is easily obtained to be:

$$\Phi(t) = \langle e^{itx} \rangle = \frac{e^{itm}\Gamma(a+itb)}{\Gamma(a)a^{itb}} \quad (16)$$

The cumulants $K_n$ are thus deduced from the successive derivatives of the logarithm of $\Phi(t)$ at t = 0:

$$\begin{cases} K_1 = \langle x \rangle = m + b(\psi(a) - \ln(a)) \quad K_2 = \langle (x - \langle x \rangle)^2 \rangle = b^2 \psi^{(1)}(a) \\ K_n = b^n \psi^{(n-1)}(a) = (-1)^{n+1} b^n n! \left[\sum_{k=0}^{\infty} \frac{1}{(a+k)^{n+1}}\right] \quad n > 1 \end{cases} \quad (17)$$

where $\psi^{(n)}(x) = \frac{d^{n+1}}{dx^{n+1}} \ln\Gamma(x)$ is a polygamma function [48]. For $a=1$, the distribution is the classical Gumbel distribution for the minimum extreme value with a mean $\langle x \rangle = m - b\gamma$, where $\gamma = 0.5772215...$ is the Euler constant.

When solely constrained to be normalized, the probability density of a GG distribution of '$u$' is:

$$\begin{cases} p_G(u) = \left[\omega_2 \, b_G^{(1+\omega_1)/\omega_2} \big/ \Gamma\big((1+\omega_1)/\omega_2\big)\right] u^{\omega_1} \exp\big(-b_G u^{\omega_2}\big) & u \geq 0 \\ \langle u^n \rangle = \Gamma\big((1+n+\omega_1)/\omega_2\big) \big/ \big(b_G^{n/\omega_2} \Gamma\big((1+\omega_1)/\omega_2\big)\big) \end{cases} \quad (18)$$

$(b_G > 0)$ then it is simply shown that the distribution of

$$x = \ln(u) \qquad (19)$$

is a generalized Gumbel distribution $g_{a,b,m}(x)$ (eq.15) with :

$$a = (1+\omega_1)/\omega_2 \;,\quad b = 1/\omega_2 \;,\quad m = b \ln\{a/b_G\} \qquad (20)$$

If the spacing 's' is distributed according to a GG distribution with $\langle s \rangle = 1$, then (eq.9):

$$b_G = \big(\Gamma(a+b)/\Gamma(a)\big)^{1/b} \qquad (21)$$

so that:

$$a = (1+\omega_1)/\omega_2 \;,\quad b = 1/\omega_2 \;,\quad m = \ln\{a^b \Gamma(a)/\Gamma(a+b)\} \qquad (22)$$

$\ln(s)$ is the electrostatic interaction energy between nearest-neigbour charges when 's' is the spacing between unprocessed eigenvalues. The Wigner surmise (eq.1) corresponds to :

$$a = (1+\beta)/2 \;,\quad b = 1/2, \; m = \ln\big(\Gamma(a)\sqrt{a}/\Gamma(a+1/2)\big) \qquad (23)$$

A Brody distribution $p_{B,\omega_B}(s)$ (eq.2) yields a Gumbel distribution:

$$a = 1, \; b = 1/(1+\omega_B) \;,\; m = -\ln\big(\Gamma\big((2+\omega_B)/(1+\omega_B)\big)\big) \qquad (24)$$

Finally, when $\omega_B \to \infty$, with $a_B = \left[\Gamma\big((2+\omega_B)/(1+\omega_B)\big)\right]^{-1} \approx 1 + \gamma/\omega_B$ and $s = 1 + z/\omega_B$ we obtain $\ln(s/a_B) \approx (z-\gamma)/\omega_B$ from which the asymptotic Brody distribution, is concluded to be a Gumbel distribution :

$$q_{B,\infty}(z) = \exp\big(z - \gamma - \exp(z-\gamma)\big) \qquad (25)$$

A Brody distribution is actually a Weibull distribution [49-50], that is one of the three fundamental distributions of suitably rescaled and shifted extreme values. This raises quite naturally the question of the distribution $p_\beta(s_{min})$ of the minimum spacing $s_{min}$.

## 5. The distribution of the minimum NN spacing of the $\beta$-HE

The notation $s_{min}$ used hereafter is restricted to minimum spacings whose average is $\langle s_{min} \rangle = 1$. Simulated distributions $p_\beta(s_{min})$, (section 3, $N=60$), are shown in figure 10. They are least-squares fitted by GG distributions (eq. 9) whose shape parameters are shown in figure 11a as a function of $\beta$ $(0 \leq \beta \leq 4)$.

For $\beta = 0$, the cumulative distribution of the minimum $x_{min}$ of $N$ iid exponential random variables $s_i, \langle s_i \rangle = 1, (i=1,..,N)$ is $F(x_{min}) = 1 - \exp(-Nx_{min})$ as that of each $s_i$ is $1 - \exp(-s_i)$. With $s_{min} = Nx_{min}$, the probability density is simply, $p_0(s_{min}) = \exp(-s_{min})$ for any $N$. It is thus a Brody distribution with $\omega_B = 0$ (eq. 2). By contrast, the distribution of the maximum spacing $x_{max} - \ln N$ is a Gumbel distribution with $a=1, b=1, m=0$ (eq.15) [49-50]. Defining $s_{max} = x_{max}/(\gamma + \ln(N))$ to get $\langle s_{max} \rangle = 1$ (eq.17), its distribution is $g_{1,b,m}(s_{max})$ with $b = -1/(\gamma + \ln(N))$ and $m = \ln N/(\gamma + \ln(N))$. The asymptotic distribution $p_\beta(s_{max})$ is found to be a Gumbel distribution in the investigated range $\beta \leq 4$.

### 5.1. The $\beta$ dependence of the distribution of the minimum NN spacing

First, $n$ successive spacings $s_k (k=1,..,n)$ are calculated between the $n+1$ eigenvalues, either raw , $\lambda_j$, or unfolded , $\lambda_j^{(u)}$, retained respectively in or from the interval $(-r, +r)$ (section 3). Then, their minimum, $d_m = \min(s_1,..,s_n) = \max(-s_1,..,-s_n)$, is determined and rescaled to get $s_{min}$. The $s_k$'s are assumed to be both identically distributed and independent. The former assumption is consistent with the stationarity of the sequence of spacings for large $n$ and with ergodicity. Indeed, the spacing distribution of the $k$th spacing is found to be independent of $k$ ($\langle s_k \rangle = 1$) and is given by $p_\beta(s)$. The latter assumption will be discussed below. The distribution of $s_{min}$ is entirely governed by the repulsion, $\propto s^\beta$, at small '$s$' which holds whichever of the set of $\beta$-H eigenvalues, raw or unfolded, is used to calculate the spacings. Near $s = 0$, the cumulative distribution function is thus :

$$F(s) = \int_0^s p_\beta(x)dx \approx K_m s^{1+\beta} \tag{26}$$

The normalization constant $K_m$ depends on the shapes of the corresponding asymptotic distributions. The distribution of $d_m$ is the known distribution of the maximum of iid variables exhibiting a power law behavior at an endpoint, here $s = 0$ ($p_\beta(s) = 0$ for $s < 0$) [50]. Then, $s_{\min} = d_{\min}(K_m N)^{1/(1+\beta)} / \Gamma((2+\beta)/(1+\beta))$, has a Weibull (Brody) distribution:

$$p_{m,\beta}(s_{\min}) = c(1+\beta)s_{\min}^\beta \times \exp\{-cs_{\min}^{1+\beta}\} \quad c = \left[\Gamma((2+\beta)/(1+\beta))\right]^{1+\beta} \tag{27}$$

The simulated distributions $p_\beta(s_{\min})$ (figure 10) agree with that conclusion when allowance is made for the matrix size dependence of the fitted shape parameters $\omega_1$ and $\omega_2$ (figure 11a). Accordingly, $\ln s_{\min}$, has a Gumbel distribution (figure 11b) because the endpoint at zero of '$s$' is rejected at $-\infty$ in that transformation.

A method to probe the above conclusion is first to simulate 'spacings' which have essentially the same distribution as those calculated from $\beta$-H eigenvalues but are independent. The Wigner surmise for various values of $\beta$ (eq.1) was then selected to simulate directly iid 'spacings', named $x_i$'s. Sequences of $n$ iid 'Wigner spacings', $x_i$ $(i=1,..,n)$, are produced by that method. Each sequence provides thus a minimum spacing $x_m = \min(x_1,..,x_n)$ and the ensemble yields the distribution $p_\beta(x_{\min})$, normalized so that $\langle x_{\min} \rangle = 1$. We deduce from the previous arguments that the asymptotic distribution of $x_m$ is a Brody distribution (eq.27). The convergence to the latter was investigated as a function of $n$ by fitting the $p_\beta(x_{\min})$'s by GG distributions (figure 12a). The parameter $\omega_1$ fluctuates little around $\beta$ as $\omega_1 = \beta$ already for a single spacing $x_1$ ($n=1$) whose distribution is just $p_{W,\beta}(x_1)$. The parameter $\omega_2$ converges approximately as $n^{-\alpha}$ to its asymptotic value $1+\beta$ with $\alpha < 1$ decreasing with $\beta$ in the studied range.

These results explain why $\omega_1$ (figure 11a) is closer to $\beta$ than $\omega_2$ is to $1+\beta$ and at least partly the origin of the deviations seen in figure 11. Part of the deviations might too be due to correlations as the successive $\beta$-H spacings are not independent.

### 5.2. The linear correlation between neighbouring spacings

The linear correlation coefficients, at $\beta$, between a central spacing, labelled $s_0$ (r.s.p. $\ln s_0$), and its neighbours $(s_1, s_2,...)$ are:

$$\rho_\beta(k) = \left[\langle s_0 s_k \rangle - \langle s_0 \rangle \langle s_k \rangle\right] / (\sigma_0 \sigma_k), \quad \sigma_k^2 = \langle s_k^2 \rangle - \langle s_k \rangle^2 \quad k = 1, 2,.. \quad (28)$$

The coefficient $\rho_\beta(k)$ was calculated for $N=50$ and $N=200$. It is essentially symmetrical ($\rho_\beta(k) \approx \rho_\beta(-k)$) and independent on $\beta$ when $\beta \gtrsim 4$. It decays rapidly with $k$ to a value which is basically zero for $k \gtrsim 4$ for any $\beta$ (and $N$ for the investigated sizes) as depicted by figure 12b. Such a rapid decrease is in agreement with the results reported in appendix N of [2] which mentions a reach of about 5 with a correlation decreasing as $\rho_\beta(k)$ : $k^{-2}$ for separated spacings ($k > 1$). The correlation coefficient between $\ln s_0$ and $\ln s_k$ is practically equal to $\rho_\beta(k)$ for any $k$ and $\beta$. Both the decrease rate and the short correlation range ensure that the distribution of $\ln s_{\min}$ converge to a Gumbel distribution [50-51] and thus that $p_\beta(s_{\min})$ tends to a Brody distribution for large $N$. However the convergence may be slow as discussed above (figure 12a). This is the reason why we restricted the range of $\beta$ in figure 11 to (0-4). Simulations of $p_\beta(s_{\min})$ are still needed for larger $\beta$ and $N$. When the temperature $1/\beta$ is close to zero, 's' becomes the difference of successive correlated Gaussian eigenvalues [30]. The correlation coefficients do not depend significantly on the temperature when $\beta \geq 1$ [33]. The low-temperature distribution of a rescaled minimum $p_\infty(s_{\min})$ is expected to be a Gumbel distribution.

### 6. Conclusion

Generalized gamma distributions with two shape parameters are excellent approximations of the asymptotic NNS distributions $p_\beta(s)$ between successive unfolded eigenvalues of matrices from the $\beta$-Hermite ensemble for any $\beta$. They account both for the level repulsion in $\sim s^\beta$ when $s \to 0$, as surmised from 2x2 $\beta$-H matrices, and for the whole shape of the $p_\beta(s)$'s where 'whole' means here in the range of '$s$' accessible to experiment or to most numerical simulations. The exact NNS distribution of the GOE is on the whole ($0 \leq s \leq \sim 4$) better fitted by $p(s) \propto s \exp(-bs^{1.886})$, still numerically simple to calculate, than it is by the Wigner surmise, $p(s) \propto s \exp(-cs^2)$. The best GG approximations to the simulated distributions coincide essentially with the Wigner surmise for $\beta > \sim 2$. The Wigner surmise might then be the exact NNS distribution when $\beta \to \infty$.

Equivalently, a generalized Gumbel distribution is concluded to be an excellent approximation of the distribution of $\ln s$ which is related to the electrostatic interaction energy between successive charges. The generalized Gumbel distribution, which is more and more frequently encountered in statistical physics, is thus appropriate too for some aspects of random matrix theory (see also [33]).

The Brody distribution, which is actually a Weibull distribution of extreme value, is the asymptotic distribution of the minimum spacing between consecutive eigenvalues, unfolded or not, of matrices from the $\beta$-Hermite ensemble as verified numerically in the present work at high temperature ($\beta \lesssim \approx 4$).

The question of the extension of the present results to other families of random matrix ensembles is open.

## Acknowledgments

We thank Prof. P.J. Forrester (University of Melbourne) for communicating some exact characteristics of the asymptotic bulk level spacing distributions and a Referee for useful suggestions.

**Appendix A : Level spacing distribution of 2x2 random matrices from the $\beta$–Hermite ensemble**

The level spacing distributions of large matrices are well approximated by those of ensembles of $N = 2$ matrices as a very close encounter of two levels is only weakly influenced by other levels (Haake [8], [36-37]). A 2x2 matrix from the $\beta$–HE is (eq. 5):

$$\mathbf{H}_{2,\beta} = \begin{bmatrix} N_1(0,1) & \chi_\beta/\sqrt{2} \\ \chi_\beta/\sqrt{2} & N_2(0,1) \end{bmatrix} \quad \text{(A-1)}$$

where $N_k(0,1)$ $(k=1,2)$ and $\chi_\beta$ are three independent random variables whose distributions are standard Gaussians and chi with $\beta$ degrees of freedom ($q_\beta(x) \propto x^{\beta-1} \exp(-x^2/2)$ $(x \geq 0)$) respectively. From eq. A-1, it is straightforward to obtain the square of the spacing $d$ between the two eigenvalues as:

$$d^2/2 = \left(N_1(0,1) - N_2(0,1)\right)^2/2 + \chi_\beta^2 \quad \text{(A-2)}$$

The two terms of the right member of eq. A-2 are independent. The first is simply the square of a $N(0,1)$ variable, that is a chi-squared variable with one degree of freedom, while the second term is a chi-squared variable with $\beta$ degrees of freedom:

$$p_{\chi^2}(x) = 2^{-\beta/2} x^{\beta/2-1} \exp(-x/2) / \Gamma(\beta/2) \qquad (x \geq 0) \qquad \text{(A-3)}$$

Its characteristic function [38]:

$$\Phi_{\chi^2}(t) = \langle e^{itx} \rangle = (1-2it)^{-\beta/2} \qquad \text{(A-4)}$$

immediately shows that the sum of two independent chi-squared variables with respective degrees of freedom $\beta$ and 1 is simply a chi-squared variable with $\beta+1$ degrees of freedom. The distribution of $y = d/\sqrt{2}$ is therefore a chi distribution with $\beta+1$ degrees of freedom:

$$q_{\beta+1}(y) \propto y^\beta \exp(-y^2/2) \qquad \text{(A-5)}$$

from which the Wigner surmise is deduced after rescaling the spacing $s = \alpha d$ so that its average becomes $\langle s \rangle = 1$. The Wigner surmise for the $\beta$–Hermite ensemble is finally:

$$p_{W,\beta}(s) = a_{W,\beta} \times s^\beta \exp(-b_{W,\beta} s^2) \qquad \text{(A-6)}$$

where $a_{W,\beta}$ and $b_{W,\beta}$ are given by eq.1. For small spacings, the repulsion between NN eigenvalues varies as $s^\beta$ (eq. A-6). It includes the classical linear, quadratic and quartic repulsion found for the GOE, the GUE and the GSE, respectively. When the temperature $1/\beta$ decreases, the eigenvalue spectrum becomes more and more rigid with charges vibrating around the zeros of the Hermite polynomials whose distribution is a Wigner semicircle [34,39]. When the latter zeros are unfolded, the resulting spectrum becomes closer and closer to a rigid picket-fence spectrum at lower and lower temperature [40]. This is reflected in the fluctuation of $s$, $\sigma_\beta^2 = \langle (s-1)^2 \rangle$, whose expansion for large $\beta$ is:

$$\sigma_\beta^2 = \frac{1}{2\beta} - \frac{3}{8\beta^2} + \frac{3}{16\beta^3} + \frac{3}{128\beta^4} + ... \qquad \text{(A-7)}$$

The rescaled spacing $(s-1)\sqrt{2\beta}$ tends to a $N(0,1)$ Gaussian for large $\beta$ as shown by figure 8 for $\beta$=1000. For that value of $\beta$, a computer simulation of 100 matrices with $N$=5000 matrices gives $\sigma^2_{1000} = 0.492 \; 10^{-3}$ while $\sigma^2_{1000} = 0.4996 \; 10^{-3}$ is calculated from eq. A-7.

**Appendix B : a double-chi ensemble of real-symmetric random matrices**

We define a $N \times N$ real-symmetric matrix from the double-chi ensemble, denoted DCHI($\beta$), as follows:

$$C_{N,\beta} = \sigma M_{N,\beta} = \sigma \begin{bmatrix} M_{11} & (d\chi_\beta)_{21}/\sqrt{2} & (d\chi_\beta)_{31}/\sqrt{2} & - & (d\chi_\beta)_{N1}/\sqrt{2} \\ (d\chi_\beta)_{21}/\sqrt{2} & M_{22} & (d\chi_\beta)_{32}/\sqrt{2} & - & - \\ (d\chi_\beta)_{31}/\sqrt{2} & (d\chi_\beta)_{32}/\sqrt{2} & - & - & - \\ - & - & - & M_{N-1,N-1} & (d\chi_\beta)_{N,N-1}/\sqrt{2} \\ (d\chi_\beta)_{N1}/\sqrt{2} & - & - & (d\chi_\beta)_{N,N-1}/\sqrt{2} & M_{NN} \end{bmatrix} \quad \text{(B-1)}$$

The $\frac{N(N+1)}{2}$ distinct matrix elements, namely $C_{kk} = \sigma M_{kk} \; (k=1,...,N)$ and $C_{ij} = \sigma M_{ij} \; (i<j, \; i=2,..,N, \; j=1,...,N)$, are independently distributed and $\sigma$ is a scale factor. Every $M_{kk}$ has a $N(0,1)$ Gauss distribution while every distinct off-diagonal element $M_{i,j}$ has a reflected chi-distribution with $\beta$ degrees of freedom, often named double-chi distribution denoted here as $d\chi_\beta$. It is defined by the density :

$$d_\beta(x) = 2^{-\beta/2} |x|^{\beta-1} \exp(-x^2/2) / \Gamma(\beta/2) \quad \text{(B-2)}$$

The $d\chi_\beta$ random variable can be simply represented as a product of two independent random variables, that of a chi with $\beta$ degrees of freedom $\chi_\beta$ with $\varepsilon$, which takes with equal probabilities the value +1 and -1. The characteristic function of a $d\chi_\beta$ random variable is easily deduced from that of $\chi_\beta$ [38] to be a confluent hypergeometric function [41]:

$$\Phi_{d\chi}(t) = \langle e^{itx} \rangle_{d\chi} = {}_1F_1\left(\frac{\beta}{2}, \frac{1}{2}; -\frac{t^2}{2}\right) \quad \text{(B-3)}$$

It tends to 1 when $\beta$ tends to zero. The ensemble DCHI(0) is thus the GDE. When $\beta$=1, the distribution $d\chi_1$ reduces to a standard normal distribution (eq. B-2). The ensemble DCHI(1) coincides thus with the GOE. When $\beta \to \infty$, the $\chi_\beta$ distribution can be written as $\sqrt{\beta} + \dfrac{X}{\sqrt{2}}$ [30], where $X$ is a $N(0,1)$ Gaussian. The ensemble DCHI($\infty$) is thus the random sign symmetric matrix investigated by Wigner [42]. It is an ensemble of real-symmetric random matrices whose diagonal elements are zero and whose distinct off-diagonal elements take at random the values $\pm\sqrt{\beta}$. Its asymptotic density of states is a semicircle [42] and it shows a linear level repulsion as does the GOE: $N \to \infty, s \to 0, p(s) \sim s$. For large $\beta$, DCHI($\infty$) is further perturbed by a GOE matrix.

A single matrix $H_{N,\beta}$ from the $\beta$-HE can be calculated from every matrix $M_{N,\beta}$ of the DCHI($\beta$) ensemble. The converse does not hold, but for $N=2$ only two matrices $M_{2,\beta} = \pm H_{2,\beta}$ (eq. B-1) are associated with a matrix $H_{2,\beta}$ (eq.5). To obtain $H_{N,\beta}$ (eq.5) it suffices to:

a) replace every off-diagonal element of the lower triangular part of $M_{N,\beta}$ by its square

b) sum up all the elements of the successive subdiagonals to get :

$$H_{k,k+1}^2 = \dfrac{1}{2}\left[\sum_{m=1}^{k}(d\chi_\beta)^2_{m+N-k,m}\right] \stackrel{d}{=} \dfrac{\chi^2_{k\beta}}{2}, \quad k=1,..,N-1 \qquad \text{(B-4)}$$

where $x \stackrel{d}{=} y$ means that the random variables $x$ and $y$ are identically distributed

c) the diagonal elements of $H_{N,\beta}$ are the $M_{kk}(k=1,..,N)$ and the elements of its first subdiagonal are the $H_{k,k+1}$ $(k=1,..,N-1)$ of eq. B-4.

By construction, the distributions of $tr(M_{N,\beta})$ and of $tr(M^2_{N,\beta})$ of $N \times N$ $DCHI(\beta)$ matrices are, for any $\beta$, identical with those of $tr(H_{N,\beta})$ and of $tr(H^2_{N,\beta})$ of the $\beta$–HE respectively. The distribution of $tr(M_{N,\beta})$ is in fact a Gaussian $N(0,N)$ and that of $tr(M^2_{N,\beta})$ is a chi-squared with $N_\beta$ degrees of freedom. The eigenvalue distribution of the ensemble of 2x2 $DCHI(\beta)$ matrices is identical with that of 2x2 $\beta$-Hermite matrices for any $\beta$ as:

$$\det\begin{bmatrix} M_{11}-\lambda & \varepsilon\chi_\beta/\sqrt{2} \\ \varepsilon\chi_\beta/\sqrt{2} & M_{22}-\lambda \end{bmatrix} = \lambda^2 - \lambda tr(H_{2,\beta}) + tr(H^2_{2,\beta}) \qquad \text{(B-5)}$$

The distribution of the spacing '$s$' between the eigenvalues of a 2x2 $DCHI(\beta)$ is thus $p_{W,\beta}(s) = a_{W,\beta} s^{\beta} \exp(-b_{W,\beta} s^2)$ (eq. A-6) but the asymptotic level repulsion remains in the universality class of the GOE for any $\beta > 0$ (figure 13). The global symmetric shape of the distribution of each element of a $N \times N\ DCHI(\beta)$ matrix, with finite moments of any order, explains that the asymptotic level repulsion results solely from the matrix symmetry, that of the GOE [42]. $M_{N,\beta}$ is a full real symmetric matrix with distinct off-diagonal elements of comparable magnitude. A transition to the Poisson distribution takes place very close to $\beta$=0 for finite $N$ ($\beta \approx 1.510^{-3}$ for $N$=40) while it is expected to occur asymptotically at $\beta$=0.

Table 1:

Shape parameter $\omega_2$ obtained by least-squares fitting the exact asymptotic distributions by GG distributions (eq.13). The exact distributions with thirty points were taken from [1] (table A14.5, page 436) and from [8] (table 4.3, p.70)

| Gaussian ensemble | $\omega_1 = \beta$ | $\omega_2$ |
|---|---|---|
| GOE | 1 | 1.886(3) |
| GUE | 2 | 1.973(2) |
| GSE | 4 | 2.007(5) |

Table 2:

Comparison of some moments, of the skewness, $\gamma_1 = \dfrac{\langle (s-\langle s \rangle)^3 \rangle}{\langle (s-\langle s \rangle)^2 \rangle^{3/2}}$, and of the coefficient of excess, $\gamma_2 = \dfrac{\langle (s-\langle s \rangle)^4 \rangle}{\langle (s-\langle s \rangle)^2 \rangle^2} - 3$, calculated from eq. 10 for $\omega_1 = 1$ and $\omega_2 = 1.886$, to the exact asymptotic values for the GOE [28] ($\beta = 1, \langle s \rangle = 1$).

| $\langle s^n \rangle$ $n$ | exact (Forrester [28], chapter 6, table 6.14) | Generalized Gamma (eq. 13) $\omega_1 = 1, \omega_2 = 1.886$ | Wigner surmise (eq.1) $\beta = 1$ |
|---|---|---|---|
| 2 | 1.28553065.. | 1.28557 | 1.27323954.. |
| 3 | 1.96143895.. | 1.96140 | 1.90985931.. |
| 4 | 3.40742133.. | 3.4070 | 3.24227787.. |
| $\gamma_1$ | 0.68718998.. | 0.68610 | 0.63111065.. |
| $\gamma_2$ | 0.37123806.. | 0.3701 | 0.24508930.. |

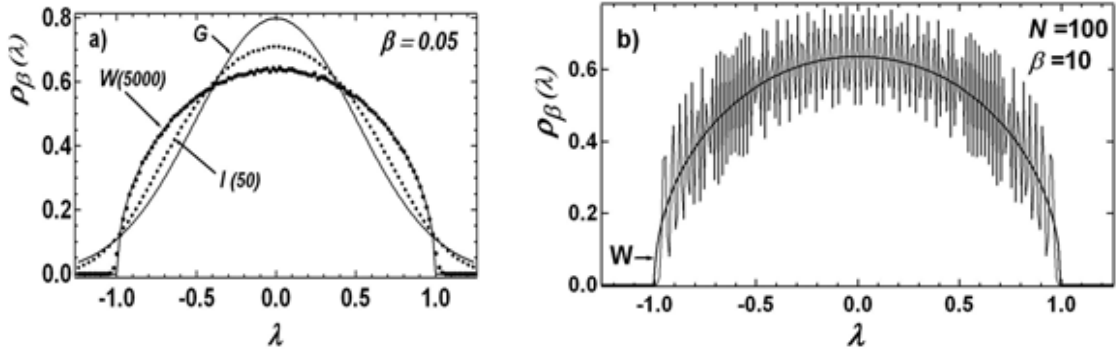

Figure 1 :

Eigenvalue density $\rho_\beta(\lambda)$ of $N \times N$ $\beta$-Hermite matrices: a) simulated for $\beta = 0.05$ with $N=5000$ and 100 matrices (dots) and with $N=50$ and $10^6$ matrices (dots, I). b) simulated for $\beta = 10$ with $N=100$ and $5.10^5$ matrices (solid line, bin size=0.01). Wigner semi-circles (eq. 7, solid lines, W) and a gaussian (solid line, G) are also are plotted. In all cases, $\langle \lambda^2 \rangle = \dfrac{1}{4}$.

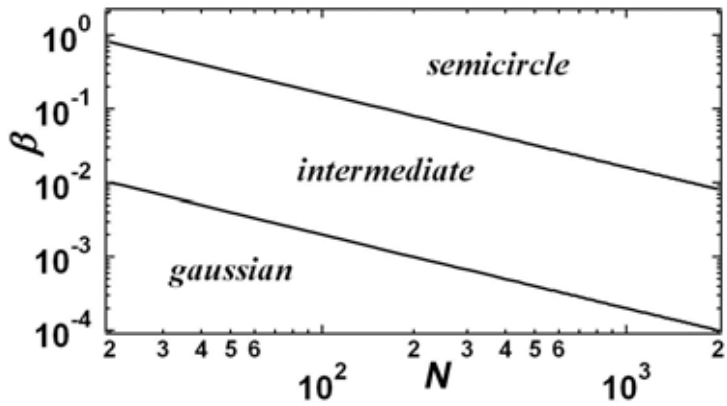

Figure 2 :

A sketch of the high-temperature evolution of the shapes of the eigenvalue densities of $N \times N$ $\beta$-H matrices, as derived from Monte Carlo simulations, as a function of $N$.

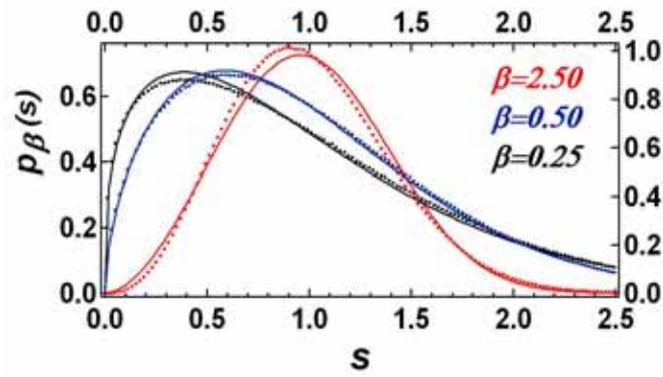

Figure 3 :

$\beta$-Hermite NNS distributions obtained from Monte Carlo simulations ($\beta$=0.25, 0.50, 2.50) and least-squares fitted by Brody distributions (eq.2) ($\beta$ increases from left to right).

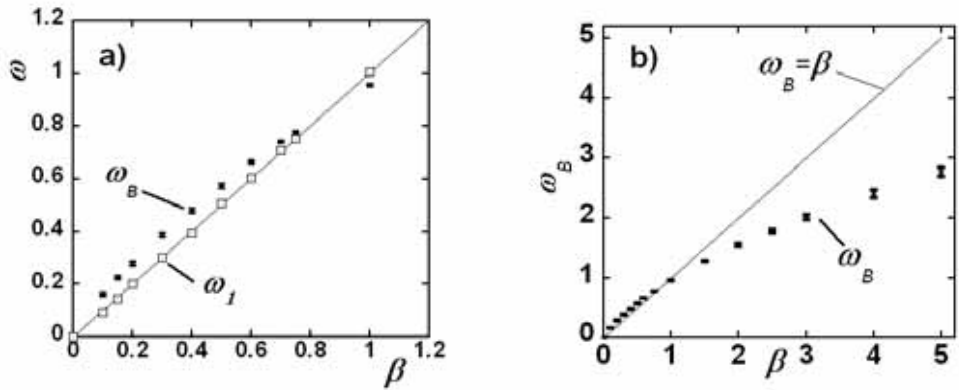

Figure 4

Shape parameter $\omega_B$ obtained from least-squares fitting of simulated $\beta$-Hermite NNS distributions by a Brody distribution (eq. 2) (0.05≤$\beta$≤5). For clarity, $\beta$ ranges: a) between 0 and 1.5 (open squares are the $\omega_1$ values of figure 5) b) between 0 and 5.

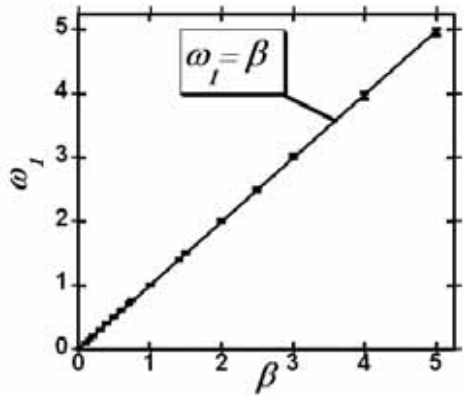

Figure 5 : Shape parameter $\omega_1$ of simulated $\beta$-Hermite NNS distributions $p_\beta(s)$ least-squares fitted by GG distributions (eq.9). The fits yield $\omega_1 = \beta$ with a very good precision in the whole range of $\beta$ (0.05≤$\beta$≤5). An expanded view of the range 0-1 is further shown in figure 4a.

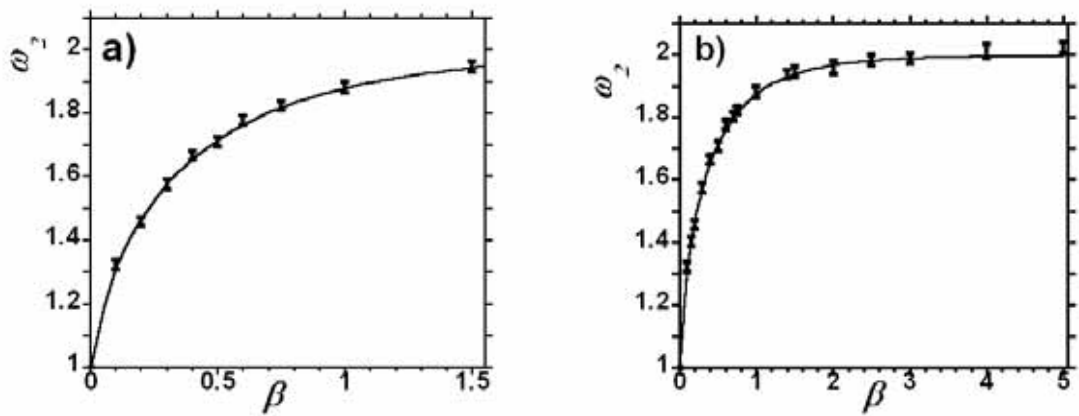

Figure 6

Shape parameter $\omega_2$ of simulated $\beta$-Hermite NNS distributions $p_\beta(s)$ least-squares fitted by GG distributions (eq.9). For clarity, $\beta$ ranges: a) between 0 and 1.5  b) between 0 and 5. The solid lines are calculated from $\omega_2 = 2 - \exp(-2.12\beta^{0.75})$.

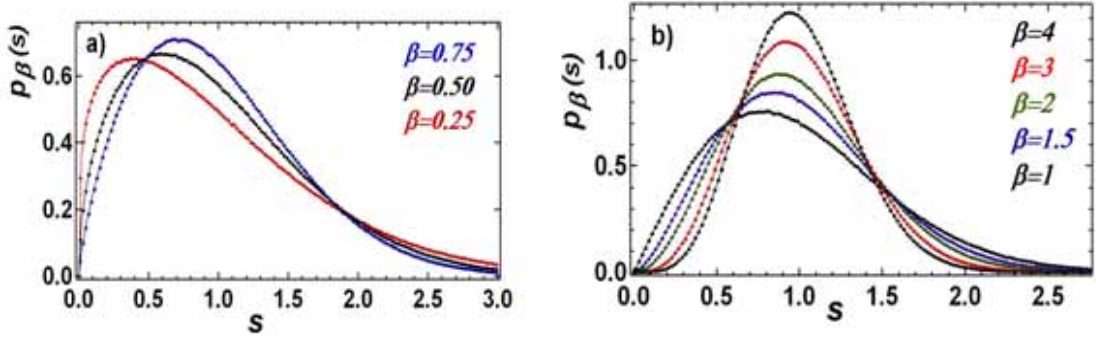

Figure 7 :

Simulated $\beta$-Hermite NNS distributions $p_\beta(s)$ as a function of $\beta$ ($1 \leq \beta \leq 4$). Solid lines are calculated from least-squares fits of the simulated data (dots) by GG distributions (eq. 13, $\omega_1 = \beta$, the sole fitted shape parameter is $\omega_2$) : a) $\beta < 1$ (the position of the maximum increases with $\beta$) b) $\beta \geq 1$ ($\beta$ increases from bottom to top for $s=1$)

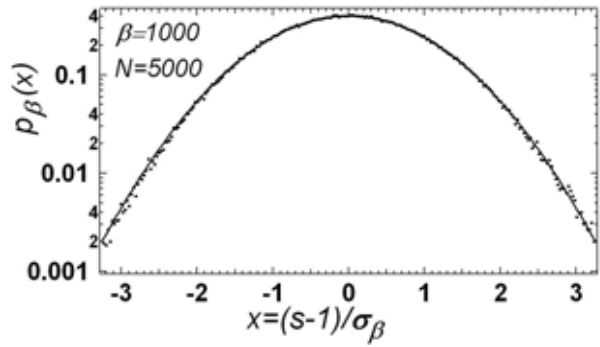

Figure 8 :

A simulated NNS distribution $p_\beta(x)$ (dots) with $x = \dfrac{s-1}{\sigma_\beta}$ for $\beta=1000$ from 100 $N \times N$ $\beta$-H random matrices with $N=5000$ (a vertical log-scale has been used to enhance the resolution in the tails, solid line = $N(0,1)$ Gaussian, $-\dfrac{\log(2\pi)+x^2}{2}$ ). The fluctuation $\sigma_\beta$, which is calculated from simulated data, is well approximated from eq. A-7.

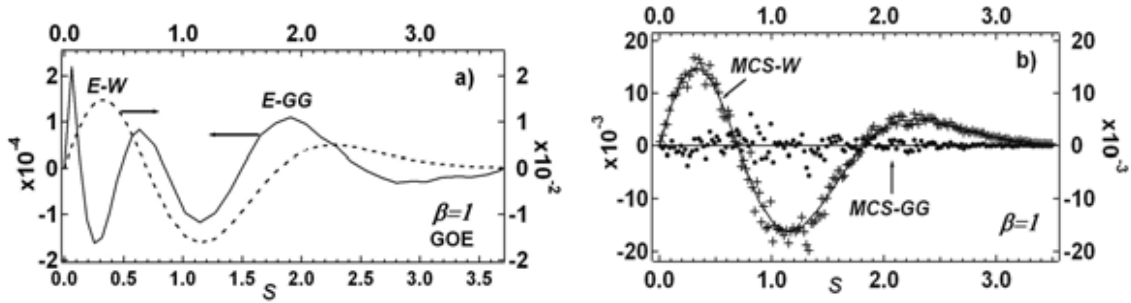

Figure 9

a) Differences between the exact asymptotic NNS distribution E of the GOE and the Wigner surmise W (eq.1, $\beta=1$) and between E and the GG distribution with $\omega_1 = 1$ and $\omega_2 = 1.886$. Note the differences between the vertical scales.

b) Differences between a distribution simulated from $10^5$ 100x100 1-Hermite matrices (MCS) and W and GG ($\omega_1 = 1$ and $\omega_2 = 1.886$) distributions; The difference E-W plotted in a) has been replotted (solid line).

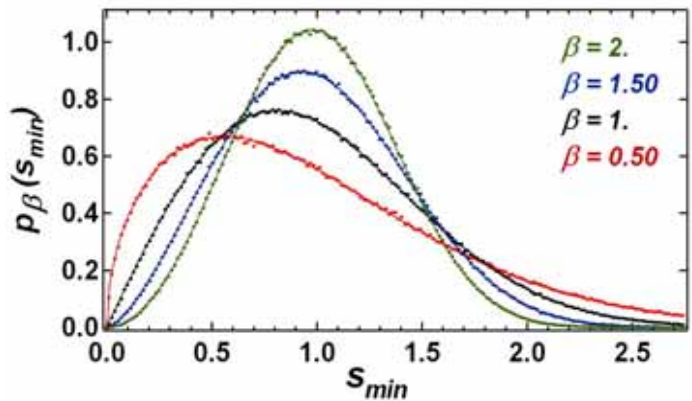

Figure 10 :

Simulated distributions of the minimum NNS $p_\beta(s_{min})$ for various values of $\beta$ (unfolded eigenvalues). Solid lines are calculated from least-squares fits of the simulated data (dots) by GG distributions $p_{\omega_1,\omega_2}(s_{min})$ (eq. 9) ($\beta$ increases from bottom to top for $s$=1).

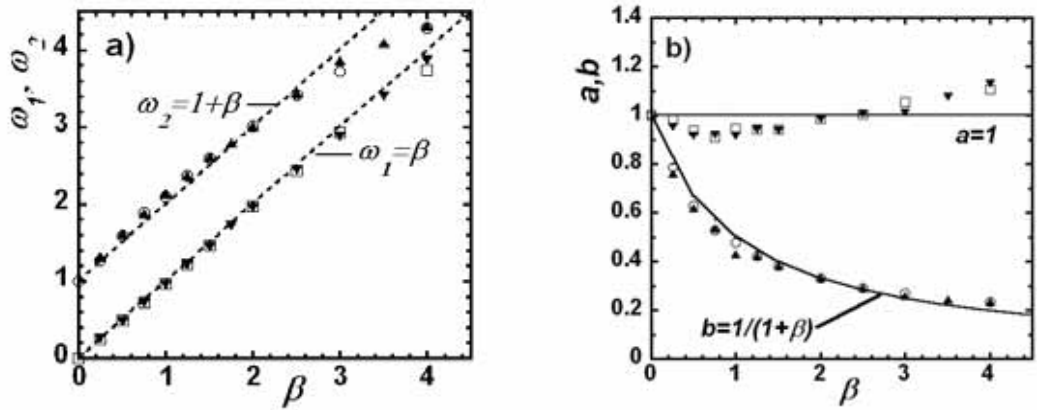

Figure 11

In a) and b) parameters represented by empty symbols and by solid triangles are associated with raw and unfolded eigenvalues respectively. a) Shape parameters $\omega_1$ and $\omega_2$ of $p_\beta(s_{\min})$ least-squares fitted by GG distributions. The dotted lines are respectively $\omega_1 = \beta$ and $\omega_2 = 1+\beta$ as expected for a Brody distribution with $\omega_B = \beta$ (eq. 2) b) parameters $a$ and $b$ of the generalized Gumbel distribution (eqs 15 and 22) as calculated from $\omega_1$ and $\omega_2$ of a). The solid line and the dotted line show the parameters of the Gumbel distribution wich is associated with a Brody distribution.

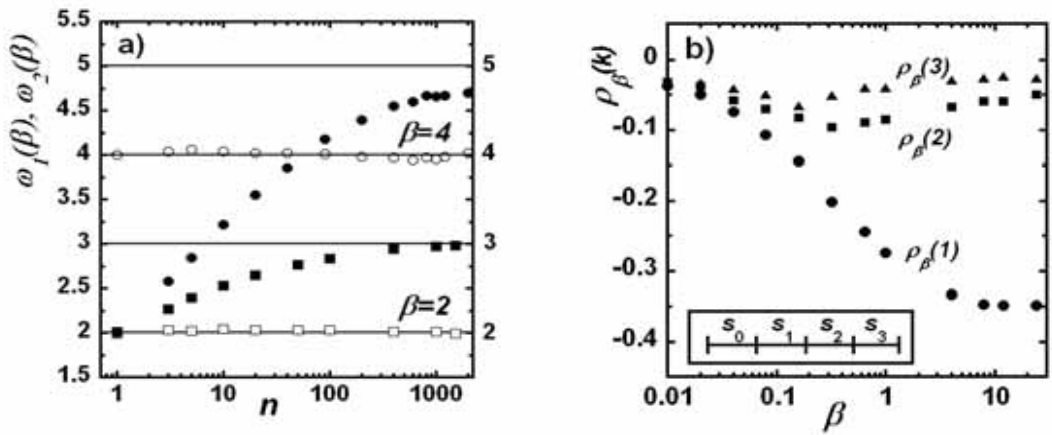

Figure 12 :

a) Shape parameters $\omega_1$ (empty symbols) and $\omega_2$ (full symbols) of the distributions of the minimum NNS of $n$ iid 'Wigner spacings' fitted by GG distributions as a function of $n$ for $\beta = 2$ (squares) and $\beta = 4$ (circles) b) Linear correlation coefficients $\rho_\beta(k)$ between a central spacing $s_0$ and its three first neighbouring spacings $s_k$ $(k = 1, 2, 3)$ (inset) as a function of $\beta$.

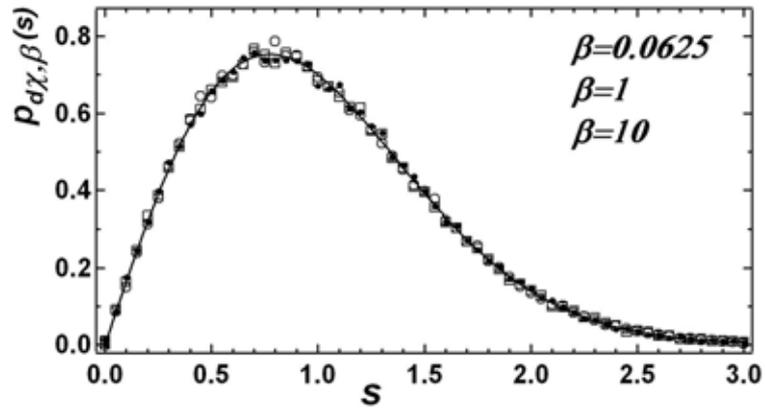

<u>Figure 13</u> :

NNS distributions $p_\beta(s)$ simulated from $10^5$ $40\times 40$ DCHI($\beta$) matrices (appendix B) ($\beta$=0.0625, empty circles, $\beta$=1, empty squares, $\beta$=10, dots). The solid line is the GG distribution of the $\text{GOE}\left(\omega_1=1, \omega_2=1.886, \text{ eq. }13\right)$.